\documentclass[showpacs,preprint]{revtex4}
\usepackage{amssymb}
\usepackage{graphicx}
\usepackage{dcolumn}
\usepackage{bm}

\begin{document}

\title{Table-top synchrotron}
\author{S.~Kiselev and A.~Pukhov }
\email{pukhov@thphy.uni-duesseldorf.de}
\affiliation{Institut fur Theoretische Physik I,
  Heinrich-Heine-Universitat Duesseldorf,
40225 Duesseldorf, Germany}
\author{I.~Kostyukov}
\affiliation{Institute of Applied Physics, Russian Academy of Science,
  46 Uljanov St.
603950 Nizhny Novgorod, Russia}
\date{\today}

\begin{abstract}
Using three-dimensional particle-in-cell simulations we show that a
strongly nonlinear plasma wave excited by an
ultrahigh intensity laser pulse works as a compact high-brightness
source of X-ray radiation. It has been recently suggested by
A.~Pukhov and J.~Meyer-ter-Vehn, Appl.~Phys.~B \textbf{74}, 355
  (2002), that in a strongly nonlinear regime the plasma wave
transforms to a ``bubble'', which is almost free from background
electrons. Inside the bubble, a dense bunch of relativistic electrons
is produced. These accelerated electrons make betatron oscillations in the
transverse fields of the bubble and emit a bright broadband
X-ray radiation with a maximum about $50$~keV. The emission is
confined to a small angle of about $0.1$~rad. In
addition, we make simulations of X-ray generation by an external
28.5-GeV electron bunch injected into the bubble. $\gamma$-quanta with
up to GeV energies are observed in the simulation in a good agreement
with analytical results. The energy conversion is efficient, leading to
a significant stopping of the electron bunch over 5~mm interaction
distance.

\end{abstract}

\maketitle

\bigskip

The development of novel high-brightness compact X-ray sources is
important for many research, industrial and medical applications.
Synchrotron light sources (SLSs) are the most intense X-ray
sources today. In an SLS, the radiation is generated as a result
of relativistic electrons scattering by a bending magnet, magnetic
undulators or wigglers \cite{bending magnet}, or by high-power
laser pulses (Compton scattering)
\cite{Leemans,Leemans1,Esarey-Thomson,Pogorelsky}. Recent
experiments, which explore the interaction of an intense 28.5-GeV
electron beam with plasma at Stanford Linear Accelerator Center
(SLAC) \cite{wang,joshi-review}, have shown that an ion channel
can be successfully used as a wiggler to produce the broadband
X-ray radiation: the electron beam propagating in plasma blows out
the background electrons and generates an ion channel.

A relativistic electron running along the ion channel undergoes
betatron oscillations about the channel axis due to the restoring
force acting on the electron by the uncompensated ion charge. For
small amplitudes, $r_{0}k_{b}\ll 1$, the betatron oscillation is
close to a harmonic motion with the betatron frequency \ $\omega
_{b}=c k_{b}=\omega _{p}/\sqrt{2\gamma } $. Here $r_{0}$ is the
radial excursion of the electron, $\omega _{p}=\sqrt{4\pi
e^{2}n_{e}/m}$ is the background plasma frequency, $n_{e}$ is the
electron density that is equal to the ion charge density in the
channel, $\gamma $ is the relativistic Lorentz factor of the
electron, $e$ is the electron charge, $m$ is the electron mass and
$c$ is the speed of light. Relativistic electrons executing
betatron oscillations in the ion channel emit short-wavelength
EM~radiation \cite{jackson,Kull}. Some features of this radiation
spectrum have been studied in the recent
publications~\cite{joshi-review,esarey1}. The fundamental
wavelength of this radiation is close to $\lambda \simeq \lambda
_{b}/\left(2\gamma ^{2}\right) $ for small-amplitude near-axis betatron
oscillations, where $\lambda _{b}=2\pi /k_{b}$. The emission at the
fundamental frequency has been call ion channel laser in the work of
D.~Whittum et al \cite{Whittum}. If the amplitude 
of the betatron oscillations becomes large, then the electron
radiates high harmonics. If the plasma wiggler strength, $K=\gamma
k_{b}r_{0}=1.33\times 10^{-10}\sqrt{\gamma n_{e}\left[
c\text{m}^{-3}\right] }r_{0}\left[ \mu \text{m}\right] $,
\noindent is so high that $K\gg 1$, then the radiation spectrum
becomes quasi-continuous broadband. It is similar to the
synchrotron spectrum, which is determined by the universal
function $S(\omega /\omega _{c})$, where $S\left( x\right)
=x\int_{x}^{\infty }K_{5/3}\left( \xi \right) d\xi $ \noindent and
$\omega _{c}$ is the critical frequency \cite{jackson}. For
frequencies well below the critical frequency $\left( \omega \ll
\omega _{c}\right) $, the spectral intensity increases with
frequency as $\omega ^{2/3}$, reaches a maximum at$\ \sim
0.29\omega _{c}$, and then drops exponentially to zero above
$\omega _{c}$. The critical frequency for a relativistic electron
in an ion channel is $\hslash \omega _{c}=(3/2)\gamma ^{3}\hslash
cr_{0}k_{b}^{2}\simeq 5\times 10^{-21}\gamma ^{2}n_{e}\left[
\text{cm}^{-3}\right] r_{0}\left[ \mu \text{m}\right] MeV.$ The
synchrotron radiation emitted from an ion channel has been
observed in a recent experiment \cite{wang}.

The synchrotron radiation is confined to a narrow angle $\theta
_{R}\simeq K/\gamma $ because of the strongly relativistic motion of the
electron. The averaged total power radiated by an electron undergoing
betatron oscillations is \cite{esarey1} $\left\langle P_{total}\right\rangle
\simeq e^{2}c\gamma ^{2}k_{p}^{4}r_{0}^{2}/12$. \noindent It follows from
this expression that the radiated power is proportional to the squared
density of ions in the channel. This fact has been confirmed in the
experiment \cite{wang}. The averaged number of photons with the mean
energy $\hslash \omega _{c}$ emitted by the electron
is $\left\langle N_{X}\right\rangle \simeq (2\pi
/9)(e^{2}/\hslash c)N_{0}K\simeq 5.6\times
10^{-3}N_{0}K$, \noindent
where $N_{0}$ is the number of betatron oscillations executed by the
electron.

In the SLAC experiment \cite{wang}, the ion channel has been produced
by the electron beam itself in the blow-out regime
\cite{blow-out_regime}, when the electron beam density, $n_{b}$, is
higher than
the plasma density. The density of a relativistic electron beam
cannot be very high because of the technology reasons. This leads
to a serious limitation on the gain in the radiated power, which
is quadratic in plasma density.

The use of a high-power laser
could overcome this limitation. The high-power laser pulse can
expel plasma electrons by its ponderomotive force and create the
ion channel \cite{laser_channel}. Moreover, the strongly nonlinear
broken-wave regime has been recently observed in 3D PIC
simulations \cite{pukhov1}. In this regime, the background
electrons are completely evacuated from the first period of the
plasma wave excited behind the laser pulse, and an
``electron bubble'' is formed.
The ion density in this bubble is many orders of magnitude higher
than that in a simple beam-plasma interaction. For example, the
ion density in the laser-produced channel
can be as high as $10^{19}$ cm$^{-3}$
\cite{pukhov1,laser_channel}. This is $10^{5}$ times higher than that
in the  recently reported beam-plasma experiment
\cite{wang}. Therefore the radiated power in the laser-produced
channel may be $10^{10}$ times higher. The bubble moves with the group
velocity of the laser pulse, which is close to the speed of light. A
relativistic electron bunch injected into the bubble can propagate inside
the bubble over a very long distance. Hence, in spite of the small length of
the bubble itself, the electrons can oscillate in the bubble for a
long time.

It has been recently shown by three-dimensional particle-in-cell (PIC)
simulations that a dense quasi-monoenergetic bunch of relativistic
electrons, collected from the background plasma, can be generated inside
the bubble \cite{pukhov1}. Because of the bubble focusing the
bunch has a much higher density than the background plasma.
In the present work we show that betatron oscillations of the bunch
in the transverse fields of the bubble lead to the efficient X-ray
generation, which can be used for the developing of table-top
high-brightness X-ray radiation sources.

We perform a numerical simulation of the X-ray generation in
laser-plasma interactions for the strongly nonlinear broken-wave
regime when the bubble is formed behind the laser pulse. We use the
fully 3D PIC code Virtual Laser-Plasma Laboratory \cite{vlpl}. The incident
laser pulse is circularly polarized, has the Gaussian envelope
$a(t,r)=a_{0}\exp (-r_{\perp }^{2}/r_{L}^{2}-t^{2}/T_{L}^{2})$,
and  the wavelength $\lambda
=0.82$~$\mu $m. Here $a = eA/mc^2$ is the relativistic laser
amplitude, $r_{L}=8.2$~$\mu $m, $T_{L}=22$~fs, $a_{0}=10$. The laser
pulse propagates in a plasma with the density $n_{e}=10^{19}$~cm$^{-3}$.

\begin{figure}
   \includegraphics[width=8cm,clip]{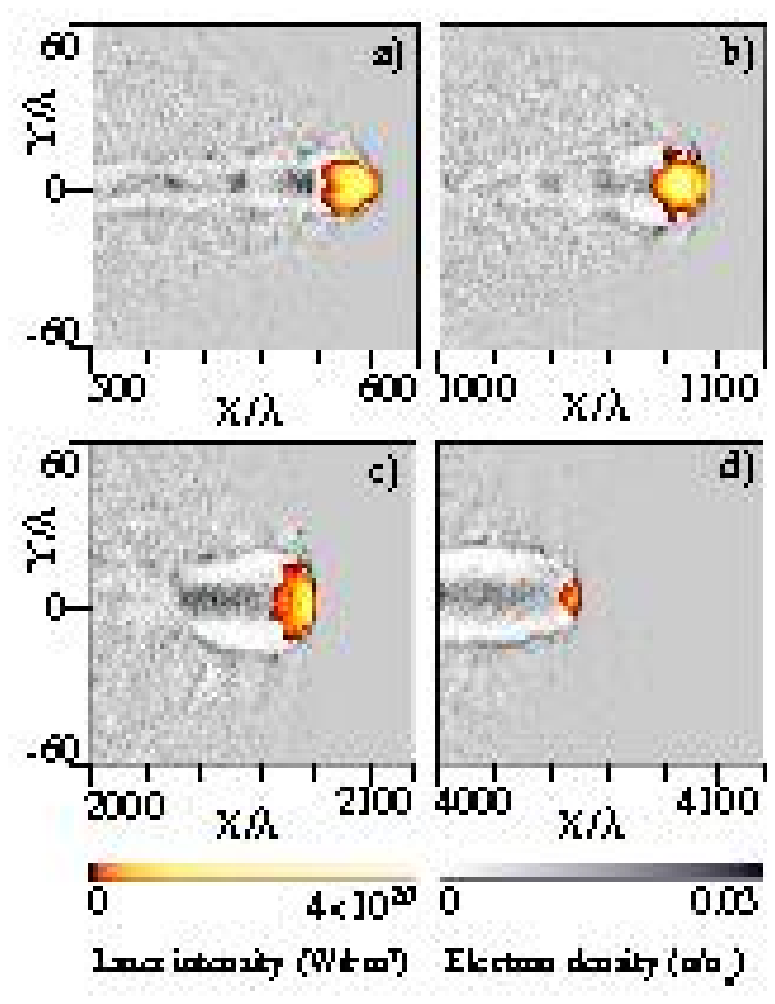}
   \vspace{9cm}
   \caption
    {
        \label{Fig1}
        (Colour) The evolution of the laser pulse intensity (the coloured
        scale) and the bubble (the electron density is given in the
        black/white scale) in the
        strongly nonlinear broken-wave regime.
        The laser pulse propagates in a plasma layer from left to
        right. The plasma density and the laser intensity
        at
        \textbf{a)} $ct/\lambda = 500,$
        \textbf{b)} $ct/\lambda = 1000,$
        \textbf{c)} $ct/\lambda = 2000,$
        \textbf{d)} $ct/\lambda = 4000.$
    }
   \end{figure}

Fig.~\ref{Fig1} presents snapshots of the laser pulse (the colored
scale) and the electron density (the black/white scale) at
different distances. The laser pulse has passed 14
Rayleigh lengths ($Z_R = \pi r^2_L/\lambda $) after the
interaction time $T_{int} = 4500 \lambda /c $. Thus, the lifetime of
the bubble is about $ 3500 \lambda /c \simeq 10$~ps. Electrons,
trapped in the bubble, form the relativistic bunch. We observe as
the bubble stretches and the bunch elongates with time.

\begin{figure}
   \includegraphics[width=8cm,clip]{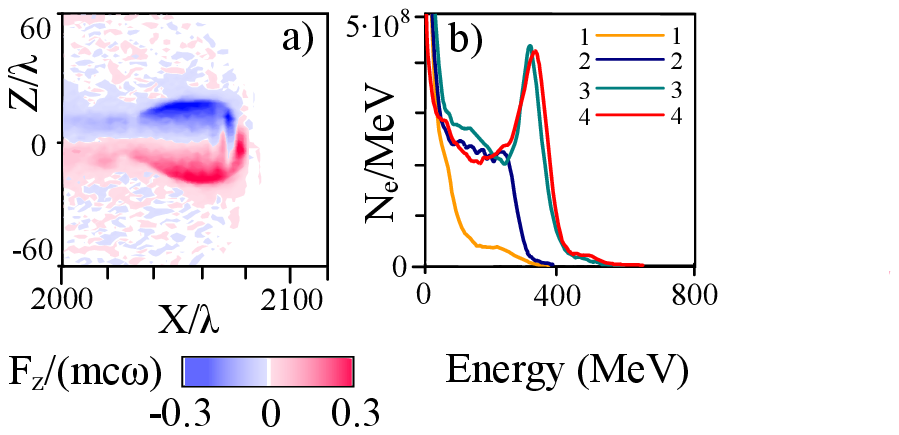}
   \vspace{12cm}
   \caption
   {
        \label{Fig2}
        (Colour) \textbf{a)} The transversal force acting on the
         relativistic electrons moving in the $x$-direction at
         $ct/\lambda = 2000$.
         \textbf{b)} Temporal variation of the energy spectrum
         of the electron bunch:
        (1) $ct/\lambda = 1000,$
        (2) $ct/\lambda = 2000,$
        (3) $ct/\lambda = 3000,$
        (4) $ct/\lambda = 4000.$
   }
   \end{figure}

Despite the fact that the bunch density is higher than the background
ion density, the transverse force acting on the accelerated electrons,
$F_{\perp }$, is mainly
determined by the electrostatic focusing force from the ions, see
Fig.~\ref{Fig2}a. This is because the charge force of relativistic
electrons and the self-generated magnetic force almost cancel each other
\cite{book-beam}. The energy spectrum of the electron bunch is
shown in Fig.~\ref{Fig2}b. We observe formation of the
quasi-monoenergetic peak \cite{pukhov1}. At $ct=4000\lambda $ the
peak is located at $360$ MeV. We calculate the corresponding wiggler
strength of $K\simeq 89\gg1$. Thus, the electrons emit X-rays in
the synchrotron regime. The number of
electrons in the bunch is about $6.5\times 10^{10}$ at this time. The
total energy of electrons of the bunch is about $3.3$
J that is about 20\% of the laser pulse energy. The number of
betatron oscillations experienced by the electrons up to this time was
$N_{0}=cT_{int}/\lambda _{b}\simeq 8.6$.

To simulate the X-ray generation we suppose that at any given 
moment of time, the electron radiation spectrum is synchrotron-like 
\cite{jackson}. The spectrum integrated over solid angle is 
defined by $S(\omega /\omega _{c})$.  The critical frequency
$\omega _{c}$ is given by the relation $\omega _{c}=(3/2)\gamma
^{2}|F_{\perp }|/(mc)$, \noindent $F_{\perp }$ is the transversal 
to the electron momentum force.
In our PIC code, we follow trajectories of each electron and
calculate the emission during the interaction.
The emitted radiation exerts a recoil on the electron
\cite{jackson}. The recoil force was included into the equations of
electron motion in our simulations.

\begin{figure}
    \includegraphics[width=8cm,height=6.2cm,clip]{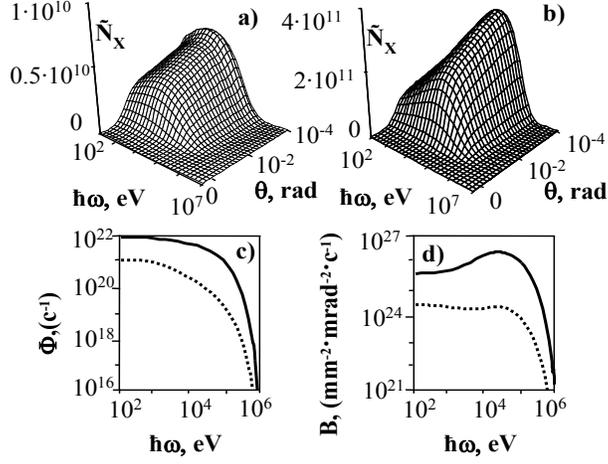}
   \vspace{12cm}
       \caption
       {           \label{Fig3}
            \textbf{a)} The synchrotron spectrum from the plasma at
            $ct/\lambda = 1000$,
            \textbf{b)} at $ct/\lambda = 4500$,
            \textbf{c)} the photon flux (the number of photons per
            second in 0.1\% bandwidth),
            \textbf{d)} the spectral brilliance.
            The dashed line in frames \textbf{c)} and \textbf{d)}
            corresponds to $ct/\lambda = 1000$, the solid line
            corresponds to $ct/\lambda = 4500.$
       }
   \end{figure}

The synchrotron spectra at\ $ct=1000\lambda $ and  $ct=4500\lambda
$ are presented in Figs.~\ref{Fig3} (a, b). The surfaces shown in
Figs.~\ref{Fig3} (a,b) give the number of photons within $0.1\%$
of the bandwidth ($\Delta \hslash \omega =10^{-3}\hslash \omega $)
per solid angle, $2\pi \sin \theta d\theta $:
$\tilde{N}_{X}=\Delta \omega d^{2}N_{X}/(2\pi \sin \theta d\omega
d\theta )$. It is seen from Fig.~\ref{Fig3} (b) that the
relativistic bunch radiates highly energetic photons within a very
narrow cone. The maximum of the radiation spectrum is located at
about $50$~keV. The analytical estimates for electron energy
predict the maximum of $S(x)$ $\simeq 0.3\hslash \omega _{c}\simeq
55$ keV that is in a good agreement with the numerical simulation
data. It is seen from Fig.~\ref{Fig3} that the radiation from the
bunch is confined within the angle $\theta \simeq 0.1$~rad and the
theoretical estimate is about $0.2$ rad. The photon flux (the
number of photons per second in $0.1\%$ bandwidth) and the spectral
brilliance of the source at $ct=1000\lambda $ and $ct=4500\lambda $
are shown in Figs.~\ref{Fig3} (c, d). We can estimate the flux and
the brilliance using the following formulas \cite{Esarey-Thomson}
$\Phi\simeq \left( \Delta \omega _{c}/\omega _{c}\right)
N_{X}(c/L_{b})$ and $B\simeq \Phi/(4\pi^{2}\theta
_{R}^{2}S_{R}^{2})\,$, where $L_{b}$ is the bunch length,
$S_{R}\simeq \pi \left[ r_{b}^{2} +c^{2}T_{int}^{2}\theta
_{R}^{2}/(4\pi ^{2})\right] $ is the effective source size of the
radiation\ and $r_{b}$ is the bunch radius.

\begin{figure}
   \includegraphics[width=8cm,clip]{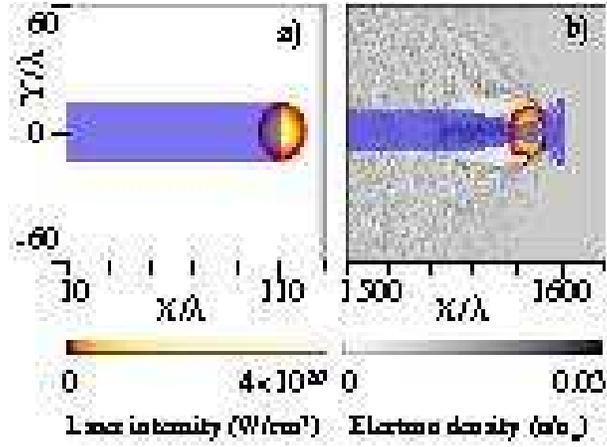}
   \vspace{12cm}
   \caption
    {
        \label{Fig4}
        (Colour) Temporal evolution of the plasma density, laser intensity
                and the envelope of the external 28.5-GeV electron bunch
                (blue): \textbf{a)} at the beginning of interaction and
                \textbf{b)} at $ct/\lambda =
                1500.$
    }
   \end{figure}

To emphasize the advantage of the X-ray generation in the
laser-produced ion channel in comparison with that in the
self-generated channel, we perform a numerical simulation of the X-ray
emission from an external 28.5-GeV electron bunch. The bunch has a
diameter $2r_{0}=24.6$~$\mu $m and a length $L_{b}=82$ $\mu $m with
the total charge $Q_{b}=5.4$~nC. The plasma and laser pulse parameters are
the same as in the previous simulation. The electron beam density was
much smaller than that of the background plasma, so that the laser pulse
and bubble dynamics is not strongly affected by the external electron
bunch. At the beginning of interaction the front of the
electron bunch is close to the center of the laser pulse (see
Figs.~\ref{Fig4} (a)). The head of the bunch has overtaken the laser
center by some $46\lambda$ after the interaction time $T_{int}=4500\lambda
/c.$ The number of betatron oscillations during the interaction time
was $N_{0}=cT_{int}/
\lambda _{b}\simeq 1.1$. It is seen from Fig.~\ref{Fig4} (b) that
the laser pulse and the bubble remain structurally stable during
the full interaction and the bunch is focused at this moment of time.

\begin{figure}
   \includegraphics[width=8cm,clip]{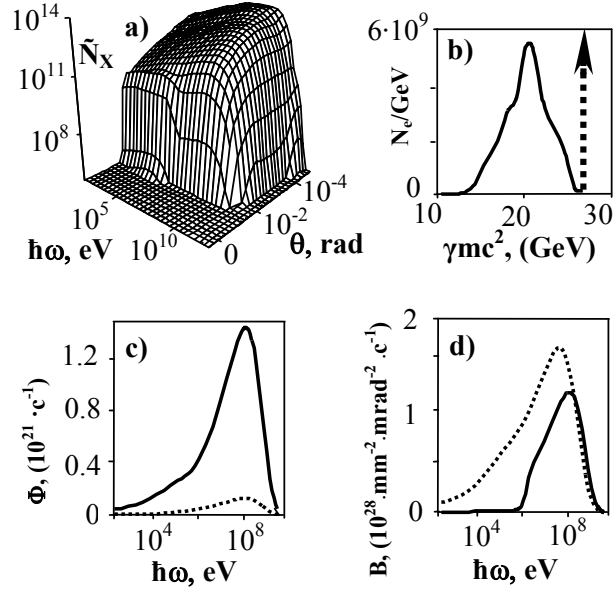}
   \vspace{10cm}
   \caption
    {
        \label{Fig5}
        \textbf{a)} Synchrotron spectrum from the external 28.5-GeV
        electron bunch at $ct/\lambda = 4500$;
        \textbf{b)} energy distribution of the bunch electrons:
        the solid line corresponds to $ct/\lambda =
                    4500,$ the dashed arrow marks the initial energy
        of the electron bunch;
        \textbf{c)} photon flux  and
        \textbf{d)} spectral brilliance.
            The dashed line in frames \textbf{c)} and
        \textbf{d)} corresponds to $ct/\lambda = 500$, the solid line
        corresponds to $ct/\lambda = 4500.$
    }
   \end{figure}

The synchrotron spectrum after the interaction time
$T_{int}=4500\lambda /c$ is presented in Fig.~\ref{Fig5} (a). In
the present simulation we do not consider the emission from the
background plasma electrons. At the given plasma density, the
plasma wiggler strength parameter is about $K \simeq 817$. It is
seen from Fig.~\ref{Fig5} (a) that the relativistic bunch radiates
highly energetic photons within a very narrow cone. The maximum of
the bunch radiation spectrum is located at about $210$~MeV. The
analytical estimates predict the maximum of $S(x)$ $\simeq
0.3\hslash \omega _{c}\simeq 385$ MeV. The disagreement is caused
by the bunch stopping because of the radiation damping force. We also
observe a significant photon flux up to the energy of $10$ GeV.
The radiation from the bunch is confined within the angle $\theta
\simeq 10$~mrad that is close to the theoretical estimate
$15$~mrad. The total number of photons emitted by the bunch are
about $2\times 10^{11}$. This means that every electron of the
bunch emits about $6$ photons. The estimation for the photon number
with the critical frequency $\omega _{c}$ is
$N_{X}=N_{e}\left\langle N_{X}\right\rangle $, where $N_{e}$ is
the number of electrons in the bunch. The estimation is in a good
agreement with the numerical simulation results. The bunch lost
about one third of its energy after $T_{int}$. The energy
distribution of the bunch electrons after the interaction is shown
in Fig.~\ref{Fig5} (b). The photon flux and the brilliance versus
the photon energy are shown in Figs.~\ref{Fig5} (c,d). The
brilliance at the beginning of interaction is slightly higher than
at the end because, at the beginning, the bunch is not yet focused
and, therefore, emits at small angles. It follows from the
Figs.~\ref{Fig5} (c,d) that the photon energy, flux and brilliance
of the X-ray emission from laser-produced ion channel are
several orders of magnitude higher than the ones observed in the
self-generated ion channel \cite{wang}.

In Conclusion, we propose a novel compact and intense x-ray radiation
source based on the strongly nonlinear broken-wave laser-plasma
interaction. The brilliance of this source is some two 
orders of magnitude higher than that of the best x-ray sources
available today \cite{TESLA}. In addition, the radiation is
polychromatic, covers that multi-keV range and comes in sub-100~fs
pulses. This bright novel source of femtosecond X-ray pulses will have 
important scientific applications by enabling the direct measurement
of atomic motion and structural dynamics in condensed matter on the
fundamental time scale of a vibrational period \cite{Leemans1}. The
100~fs  time scale is characteristic for atomic motion associated with
ultrafast chemical reactions, nonequilibrium phase transitions,
surface dynamics and even ultrafast biological processes. 

In addition, our proposed radiation source provides a sufficient
number of photons per pulse to carry out these studies in a
single-shot regime. This is crucially important, e.g., for the
biological processes. The polychromaticity of the radiation source may
allow to probe simultaneously different atomic species in complex or
disordered materials. 

The proposed radiation source can be a table-top
laser, and the plasma interaction length is less than a
centimeter. This must be compared with many 100~m structures of the 
conventional synchrotron sources \cite{TESLA}. The high ion density in the
laser-plasma wiggler provides several orders of magnitudes higher
energies of the x-ray photons than that observed in the
recent experiment with self-generated ion channels \cite{wang} and in
the designed FELs. 

One of the authors (I.~K.) gratefully acknowledges the hospitality of the
Institute for Theoretical Physics of Duesseldorf University. This work has
been supported by the Alexander von Humboldt Foundation, DFG and BMBF
(Germany).

\newpage

\newpage

\newpage

\newpage

\newpage

\end{document}